\documentclass[aps,twocolumn,prd,showpacs,showkeys,preprintnumbers,superscriptaddress,nobibnotes,floatfix,longbibliography]{revtex4-1}

\pdfoutput=1
\usepackage{amsmath}
\usepackage{amsfonts}
\usepackage{amssymb}
\usepackage{mathrsfs}
\usepackage{graphicx}
\usepackage{color}
\usepackage{longtable}
\usepackage{hyperref}
\usepackage{multirow}

\newcommand{\ie}{{\it i.e.}}

\newcommand{\eg}{{\it e.g.}}

\newcommand{\eq}{Eq.}

\newcommand{\fig}{Fig.}
\newcommand{\Fig}{Fig.}

\newcommand{\Ref}{Ref.}
\newcommand{\Refs}{Refs.}

\newcommand\brabar{\raisebox{-2.0pt}{\scalebox{.2}{ \textbf{(}}}\raisebox{-2.0pt}{{\_}}\raisebox{-2.0pt}{\scalebox{.2}{\textbf{ )}}}}

\newcommand{\equ}[1]{\eq~(\ref{equ:#1})}
\newcommand{\figu}[1]{\fig~\ref{fig:#1}}

\newcommand{\bi}{\begin{itemize}}
\newcommand{\ei}{\end{itemize}}

\begin{document}

\title{Extracting the Energy-Dependent Neutrino-Nucleon Cross Section\\Above 10~TeV Using IceCube Showers}

\author{Mauricio Bustamante}
\email{mbustamante@nbi.ku.dk}
\thanks{ORCID: \href{http://orcid.org/0000-0001-6923-0865}{0000-0001-6923-0865}}
\affiliation{Niels Bohr International Academy \& Discovery Center, Niels Bohr Institute, Blegdamsvej 17, 2100 Copenhagen, Denmark}
\affiliation{Center for Cosmology and AstroParticle Physics (CCAPP), The Ohio State University,
        Columbus, OH 43210, USA}
\affiliation{Department of Physics, The Ohio State University, Columbus, OH 43210, USA}
\author{Amy Connolly}
\email{connolly@physics.osu.edu}
\thanks{ORCID: \href{http://orcid.org/0000-0003-0049-5448}{0000-0003-0049-5448}}
\affiliation{Center for Cosmology and AstroParticle Physics (CCAPP), The Ohio State University,
        Columbus, OH 43210, USA}
\affiliation{Department of Physics, The Ohio State University, Columbus, OH 43210, USA}

\date{January 12, 2019}

\begin{abstract}

Neutrinos are key to probing the deep structure of matter and the high-energy Universe.  
Yet, until recently, their interactions had only been measured at laboratory energies up to about 350~GeV.
An opportunity to measure their interactions at higher energies opened up with the detection of high-energy neutrinos in IceCube, partially of astrophysical origin.
Scattering off matter inside the Earth affects the distribution of their arrival directions --- from this, we extract the neutrino-nucleon cross section at energies from 18 TeV to 2 PeV, in four energy bins, in spite of uncertainties in the neutrino flux.  
Using six years of public IceCube High-Energy Starting Events, we explicitly show for the first time that the energy dependence of the cross section above 18 TeV agrees with the predicted softer-than-linear dependence, and reaffirm the absence of new physics that would make the cross section rise sharply, up to a center-of-mass energy $\sqrt{s} \approx 1$ TeV.

\end{abstract}


\maketitle


{\bf Introduction.---}  Neutrino interactions, though feeble, are important for particle physics and astrophysics. 
They provide precise tests of the Standard Model\ \cite{Brock:1993sz, Conrad:1997ne, Formaggio:2013kya}, probes of new physics\ \cite{Connolly:2011vc, Chen:2013dza, Marfatia:2015hva}, and windows to otherwise veiled regions of the Universe.  Yet, at neutrino energies above 350 GeV there had been no measurement of their interactions.  This changed recently when the IceCube Collaboration found that the neutrino-nucleon cross section from 6.3 to 980 TeV agrees with predictions\ \cite{Aartsen:2017kpd}.

\begin{figure}[t!]
 \centering
 \includegraphics[width=\columnwidth]{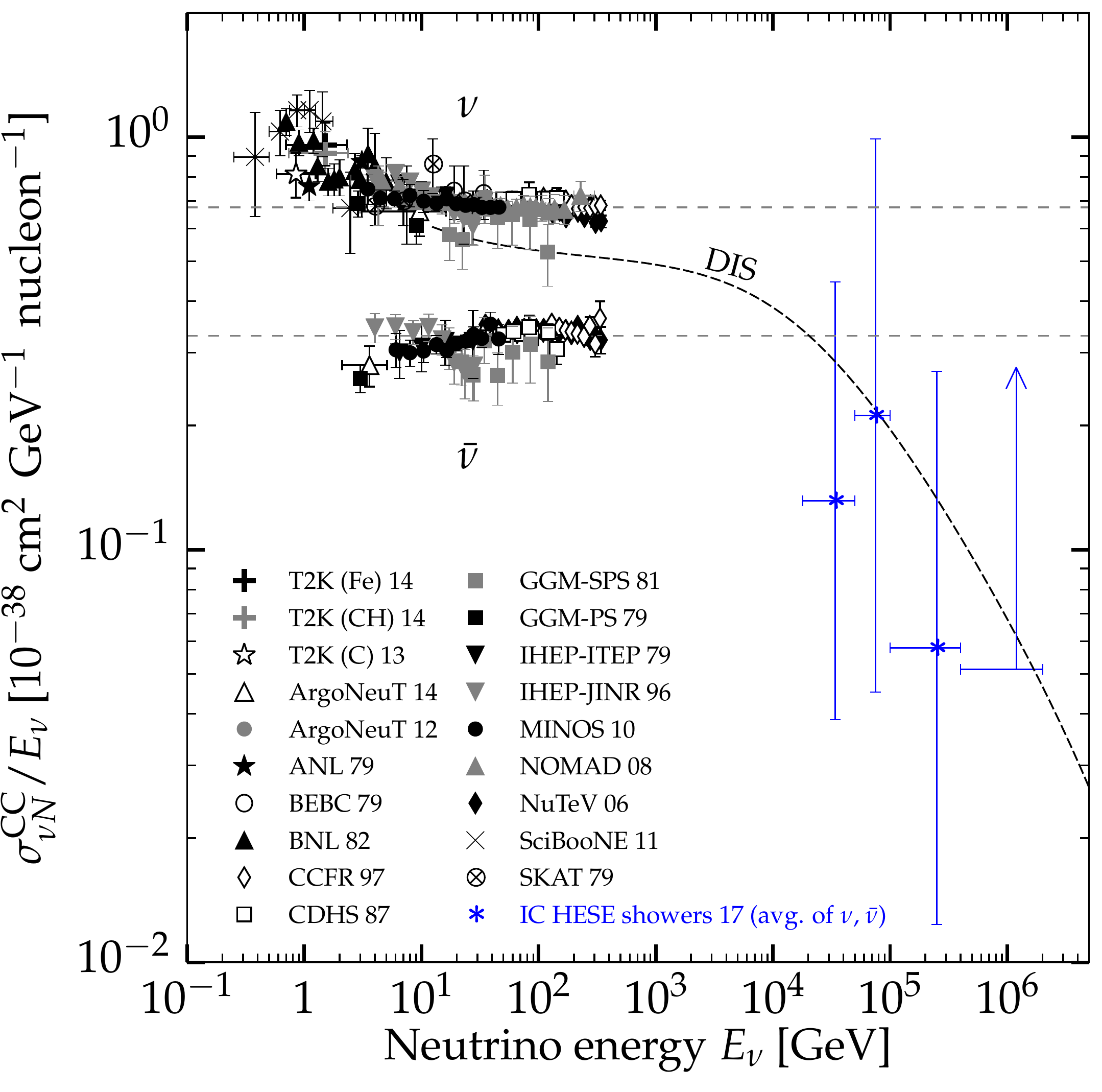}
 \caption{\label{fig:cross-section}Charged-current inclusive neutrino-nucleon cross section measurements\ 
 \cite{Mukhin:1979bd, Baranov:1978sx, Barish:1978pj, Ciampolillo:1979wp, deGroot:1978feq, Colley:1979rt, Morfin:1981kg, Baker:1982ty, Berge:1987zw, Anikeev:1995dj, Seligman:1997, Tzanov:2005kr, Wu:2007ab, Adamson:2009ju, Nakajima:2010fp, Abe:2013jth, Acciarri:2014isz, Abe:2014nox}.  The new results from this work, based on 6 years of IceCube HESE showers\ \cite{Aartsen:2014gkd,Kopper:2015vzf,IC4yrHESEURL,TalkKopperICRC2017}, are an average between cross sections for $\nu$ and $\bar{\nu}$, assuming equal astrophysical fluxes of each.  In the highest-energy bin, we only set a lower limit ($1\sigma$ shown).  The thick dashed curve is a standard prediction of deep inelastic scattering (DIS), averaged between $\nu$ and $\bar{\nu}$.  Horizontal thin dashed lines are global averages from \Ref\ \cite{Olive:2016xmw}, which do not include the new results.}
\end{figure}

Because there is no artificial neutrino beam at a TeV and above, IceCube used atmospheric and astrophysical neutrinos, the latter discovered by them up to a few PeV\ \cite{Aartsen:2013bka,Aartsen:2013jdh,Aartsen:2013eka,Aartsen:2014gkd,Aartsen:2014muf,Aartsen:2015xup,Aartsen:2015knd,Aartsen:2015rwa,Aartsen:2016xlq}.  \Refs\ \cite{Hooper:2002yq, Hussain:2006wg, Borriello:2007cs, Hussain:2007ba, Connolly:2011vc, Marfatia:2015hva} showed that, because IceCube neutrinos interact significantly with matter inside Earth, their distribution in energy and arrival direction carries information about neutrino-nucleon cross sections, which, like IceCube\ \cite{Aartsen:2017kpd}, we extract.

However, \Ref\ \cite{Aartsen:2017kpd} extracted the cross section in a single, wide energy bin, so its energy dependence in that range remains untested.  A significant deviation from the predicted softer-than-linear dependence could signal the presence of new physics, so we extract the cross section in intervals from 18~TeV to 2~PeV.  While \Ref\ \cite{Aartsen:2017kpd} used only events born outside of IceCube we use instead only events born inside of it, which leads to a better handle on the neutrino energy.

Figure \ref{fig:cross-section} shows that the cross section that we extract is compatible with the standard prediction.  There is no indication of the sharp rise, at least below 1 PeV, predicted by some models of new physics\ \cite{AlvarezMuniz:2001mk, AlvarezMuniz:2002ga, Anchordoqui:2003jr, Ahn:2003cza, Illana:2004qc, Hussain:2005dm, Anchordoqui:2006fn, Lykken:2007kp, Anchordoqui:2010hq, Marfatia:2015hva}.


{\bf Neutrino-nucleon cross section.---}  Above a few GeV, neutrino-nucleon interactions are typically deep inelastic scatterings (DIS), where the neutrino scatters off one of the constituent partons of the nucleon --- a quark or a gluon.  In both the charged-current (CC, $\overset{\brabar}{\nu}_l + N \to l^\mp + X$) and neutral-current (NC, $\overset{\brabar}{\nu}_l + N \to \overset{\brabar}{\nu}_l + X$) forms of this interaction, the nucleon $N$ is broken up into partons that hadronize into a final state $X$.  
The final-state hadrons carry a fraction $y$ --- the inelasticity --- of the initial neutrino energy, while the final-state lepton carries the remaining fraction $(1-y)$.

Calculation of the cross section $\sigma_{\nu N}$ requires knowing the parton distribution functions (PDFs) in the nucleon.  PDFs depend on two kinematic variables: $Q^2 \equiv -q^2$, the four-momentum transferred to the mediating $W$ or $Z$ boson, and the Bjorken scaling $x$, the fraction of nucleon momentum carried by the interacting parton\ \cite{Giunti:2007ry}

\begin{figure}[t]
 \centering
 \includegraphics[width=\columnwidth]{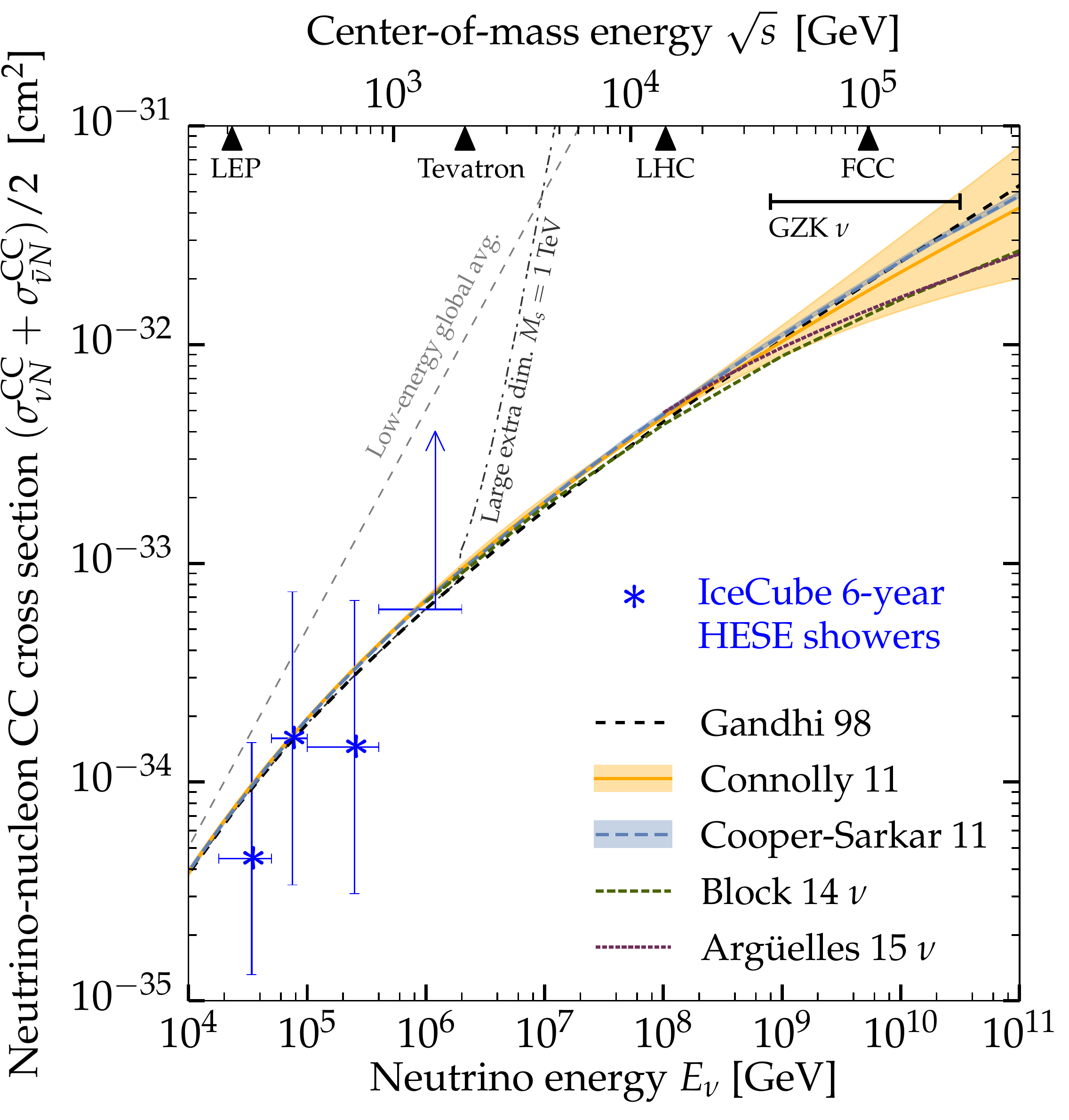}
 \caption{\label{fig:cross_sections_compare}Neutrino-nucleon charged-current cross section, averaged for neutrinos and anti-neutrinos, from different predictions (lines) \cite{Gandhi:1998ri, Connolly:2011vc, CooperSarkar:2011pa, Block:2014kza, Arguelles:2015wba}, compared to measurements from this work (stars).  The low-energy global average\ \cite{Olive:2016xmw} has the linear dependence on $E_\nu$ appropriate below $\sim$10 TeV.  The model of large extra dimensions, included for illustration, is from \Ref\ \cite{AlvarezMuniz:2001mk} (quantum-gravity scale of 1 TeV and all partial waves summed), corrected here to match modern standard predictions of the cross section below 1 PeV.}
\end{figure}

To compute cross sections at neutrino energies $E_\nu$ between TeV and PeV, we need PDFs evaluated at $x \gtrsim m_W / E_\nu \sim 10^{-4}$.  Because these are known --- at low $x$, from $ep$ collisions in HERA\ \cite{Aaron:2009aa, Abramowicz:2015mha} --- the uncertainty in the predicted TeV--PeV cross sections is small.  \Refs\ \cite{Gandhi:1995tf, Gandhi:1998ri, CooperSarkar:2007cv, Gluck:2010rw, Connolly:2011vc, CooperSarkar:2011pa, Block:2014kza, Goncalves:2014woa, Arguelles:2015wba, Albacete:2015zra, Gauld:2016kpd, Ball:2017otu} have performed such calculations, some of which are shown in \figu{cross_sections_compare}.  Below $\sim$10 TeV, they yield $\sigma_{\nu N} \propto E_\nu$, revelatory of hard scattering off partons, and in agreement with data.  Above $\sim$10 TeV, where $Q^2 \sim m_W^2$, they yield a softer-than-linear energy dependence, which has only been glimpsed in the available data up to 350 GeV\ \cite{Brock:1993sz, Conrad:1997ne, Formaggio:2013kya}.


{\bf Detecting high-energy neutrinos.---}  IceCube is the largest optical-Cherenkov neutrino detector.  It consists of strings of photomultipliers buried deep in the clear Antarctic ice, instrumenting a volume of about 1 km$^3$.

\begin{figure}[t]
 \centering
 \includegraphics[width=\columnwidth]{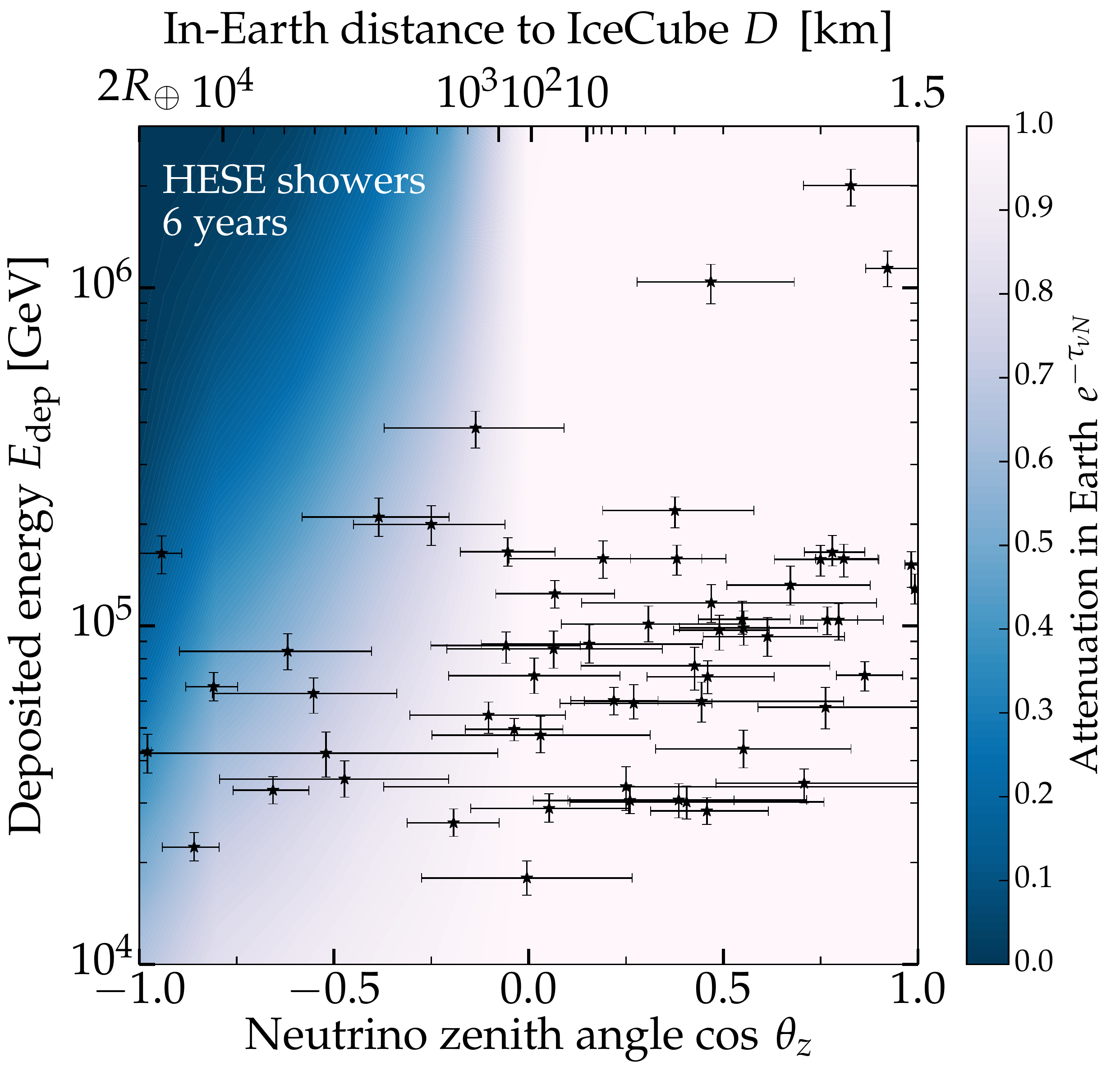}
 \caption{\label{fig:hese-costhz-energy}Neutrino-induced showers from the IceCube 6-year HESE\ \cite{Aartsen:2014gkd,Kopper:2015vzf,IC4yrHESEURL,TalkKopperICRC2017} sample.  Neutrinos arrive from above ($\cos \theta_z > 0$); from below, through the Earth ($\cos \theta_z < 0$); and horizontally ($\cos \theta_z = 0$).  They travel a distance $D$ inside the Earth (of radius $R_\oplus = 6371$ km) to IceCube, buried at a depth of 1.5 km.  The background shading represents the fraction of isotropic neutrino flux that survives after being attenuated by $\nu N$ interactions inside the Earth, calculated using cross sections predicted in \Ref\ \cite{CooperSarkar:2011pa}.}
\end{figure}

Above TeV, CC interactions of $\nu_e$ and $\nu_\tau$ with nucleons in the ice, and NC interactions of all flavors, create localized particle showers, with roughly spherical Cherenkov-light profiles centered on the interaction vertex.  CC interactions of $\nu_\mu$ additionally create muons that make elongated tracks of Cherenkov light, several kilometers long and easily identifiable.  (Other, flavor-specific signatures require energies higher than in our analysis\ \cite{Glashow:1960zz, Learned:1994wg, Beacom:2003nh, Bugaev:2003sw, Anchordoqui:2004eb, Bhattacharya:2011qu, Bhattacharya:2012fh, Barger:2014iua, Palladino:2015uoa}.) 

From the amount of collected light in a detected event, and its spatial and temporal profiles, IceCube infers its energy and arrival direction.  But it cannot distinguish neutrinos from anti-neutrinos, or NC from CC showers, since they make similar light signals.


{\bf Using contained showers only.---}  Because cross sections vary with neutrino energy, we use exclusively a class of IceCube events where the incoming neutrino energy can be inferred using as few assumptions as possible.
These are ``starting events'', where the neutrino interaction was contained in the detector.  
Of these, we use only showers, not tracks, due not to a fundamental limitation, but to the IceCube data that is publicly available.
Our approach differs from that of \Ref\ \cite{Aartsen:2017kpd}, which used only through-going muons, born in neutrino interactions outside the detector, for which estimation of the neutrino energy requires making important assumptions.

We use the publicly available 6-year sample of IceCube High Energy Starting Events (HESE)\ \cite{Aartsen:2014gkd,Kopper:2015vzf,IC4yrHESEURL,TalkKopperICRC2017}, consisting of 58 contained showers with deposited energies $E_\text{dep}$ from 18 TeV to 2 PeV.  Below a few tens of TeV, about half of the showers is due to atmospheric neutrinos and half to astrophysical neutrinos\ \cite{TalkKopperICRC2017}; above, showers from astrophysical neutrinos dominate\ \cite{Beacom:2004jb, Laha:2013eev}.

Figure \ref{fig:hese-costhz-energy} shows the HESE showers distributed in $E_\text{dep}$ and zenith angle $\theta_z$.  Representative uncertainties are 10\% in $E_\text{dep}$ and $15^\circ$ in $\theta_z$\ \cite{Aartsen:2013vja}, which we adopt to describe the detector resolution.  Showers are scarce above 200 TeV because the neutrino flux falls steeply with $E_\nu$.

In CC showers, the full neutrino energy is deposited in the ice, \ie, $E_\text{dep} \approx E_\nu$, because both the outgoing electron or tau and the final-state hadrons shower.  In NC showers, only a fraction $y$ of the neutrino energy is deposited in the ice, \ie, $E_\text{dep} = y E_\nu$, because only final-state hadrons shower.  Standard calculations yield an average $\langle y \rangle = 0.35$ at 10 TeV and 0.25 at 1 PeV\ \cite{Gandhi:1995tf}.  Because of this low value and because the neutrino fluxes fall steeply with $E_\nu$,
NC showers are nominally sub-dominant at any value of $E_\text{dep}$.  

In starting tracks, the shower made by final-state hadrons is contained by the detector, but the muon track typically exits it.  An assumption-free reconstruction of $E_\nu$ requires knowing separately the energy of the hadronic shower $E_\text{sh}$ and the muon energy loss rate $dE_\mu/dX$, which is proportional to the muon energy $E_\mu$\ \cite{Aartsen:2013vja}.  Yet, while these quantities are known internally to the IceCube Collaboration, public data only provides, for each starting track, the total deposited energy, $E_\text{sh} + \lvert dE_\mu/dX \rvert \Delta X$, where $\Delta X$ is the track length in the detector.  Without additional information, in order to deduce $E_\nu$, we would need to assume values of $y$ and $\Delta X$ for each event\ \cite{Palomares-Ruiz:2015mka}.  Hence, in keeping to our tenet of using few assumptions to deduce $E_\nu$, we do not include starting tracks in our analysis.  This choice also reduces the chance of erroneously using a track created by an atmospheric muon, not a neutrino.  

To use through-going muons in extracting the cross section, IceCube\ \cite{Aartsen:2017kpd} inferred the most likely parent neutrino energy from the measured muon energy\ \cite{Aartsen:2013vja} by assuming the inelasticity distribution $d\sigma_{\nu N} / dy$\ from \Ref\ \cite{CooperSarkar:2011pa}.
By using only contained showers, we forgo the need to assume an inelasticity distribution, and remain more sensitive to potential new physics that could modify it.


{\bf Sensitivity to the cross section.---}  Neutrino-nucleon interactions make the Earth opaque to neutrinos above 10 TeV, so neutrino fluxes are attenuated upon reaching IceCube.  More neutrinos reach it from above --- after crossing a few kilometers of ice --- than from below --- after crossing up to the diameter of the Earth.  

A flux of incoming neutrinos with energy $E_\nu$ and zenith angle $\theta_z$ is attenuated by a factor $e^{-\tau_{\nu N}(E_\nu,\theta_z)} \equiv \exp \left[ - D(\theta_z) / {L_{\nu N}}(E_\nu,\theta_z)  \right]$, where $\tau_{\nu N}$ is the opacity to $\nu N$ interactions, $D$ is the distance from the point of entry into Earth to IceCube, and $L_{\nu N} = m_N / [ (\sigma_{\nu N}^{\text{CC}}(E_\nu) + \sigma_{\nu N}^{\text{NC}}(E_\nu) ) \langle \rho_\oplus(\theta_z) \rangle ]$ is the neutrino interaction length.  Here, $\sigma_{\nu N}^{\text{CC}}$ and $\sigma_{\nu N}^{\text{NC}}$ are, respectively, the CC and NC cross sections, $m_N$ is the average nucleon mass in isoscalar matter, and $\langle \rho_\oplus \rangle$ is the average matter density along this direction, calculated using the density profile from the Preliminary Reference Earth Model\ \cite{Dziewonski:1981xy,Gandhi:1995tf}.  Details are in the Supplemental Material, which includes \Refs\ \cite{Feroz:2007kg, Kelner:2008ke, Feroz:2008xx, Hummer:2010vx, Aartsen:2013rt, Feroz:2013hea, Robitaille:2013mpa, Buchner:2014nha, Blum:2014ewa, Bustamante:2016ciw, Vincent:2017svp, IC2yrMESEURL, KleinPrivateComm, TalkXuICRC2017}.  Attenuation grows with the cross sections --- which grow with $E_\nu$ --- and with $D$; both effects are evident in the background shading in \figu{hese-costhz-energy}.

Within an energy interval, the number of events induced by a neutrino flux $\Phi_\nu$ is $N_\text{sh} \propto \Phi_\nu \cdot e^{-\tau_{\nu N}} \cdot \sigma_{\nu N}$.  Downgoing showers ($\cos \theta_z > 0$) --- unaffected by attenuation --- fix the product $\Phi_\nu \cdot \sigma_{\nu N}$, while upgoing showers ($\cos \theta_z < 0$) --- affected by attenuation --- break the degeneracy between $\Phi_\nu$ and $\sigma_{\nu N}$ via $e^{-\tau_{\nu N}}$, providing sensitivity to the cross sections.


{\bf Extracting cross sections.---}  We propagate atmospheric and astrophysical neutrinos through the Earth and produce test samples of contained showers in IceCube, taking into account its energy and angular resolution; see the Supplemental Material.  To extract the cross sections, we compare the distributions in $E_\text{dep}$ and $\cos \theta_z$ of the test showers --- generated with varying values of the cross sections --- to the distribution observed by IceCube.

To probe the energy dependence of the cross sections, we bin showers in $E_\text{dep}$ and extract the cross section from data in each bin independently of the others.   Except for global assumptions on detector resolution and the choice of atmospheric neutrino spectrum (see below), parameters extracted in different bins are uncorrelated.

The first three bins contain comparable numbers of showers: 18--50 TeV (17 showers), 50--100 TeV (18 showers), and 100--400 TeV (20 showers).  The final bin, 400--2004 TeV, contains only 3 downgoing showers, between 1--2 PeV.  Due to their short travel distances ($D \lesssim 10$ km) and negligible expected attenuation, in this bin we only set a lower limit on the cross section.   This stresses the need for upgoing HESE events above 400 TeV.

For atmospheric neutrinos, we use the most recent calculation of the $\nu_e$, $\bar{\nu}_e$, $\nu_\mu$, and $\bar{\nu}_\mu$ fluxes from pion and kaon decays from \Ref\ \cite{Honda:2015fha}.  Their zenith-angle distribution at the South Pole, though anisotropic, is symmetric around $\cos \theta_z = 0$, so it does not introduce spurious directional asymmetries.
We do not include a contribution from prompt atmospheric neutrinos\ \cite{Bugaev:1989we, Gondolo:1995fq, Pasquali:1998ji, Enberg:2008te, Gaisser:2013ira, Bhattacharya:2015jpa, Garzelli:2015psa, Gauld:2015kvh, Fedynitch:2015zbe, Halzen:2016pwl, Halzen:2016thi, Gaisser:2016obt,  Bhattacharya:2016jce, Engel:2017wpa}, since searches have failed to find evidence of them\ \cite{Aartsen:2013bka, Aartsen:2013jdh, Aartsen:2013eka, Aartsen:2014gkd, Aartsen:2014muf, Aartsen:2015xup, Aartsen:2015knd, Aartsen:2015rwa, Aartsen:2016xlq}.  We include the self-veto\ \cite{Schonert:2008is, Gaisser:2014bja} used by the HESE analysis to reduce the atmospheric contribution.

For astrophysical neutrinos, we assume, independently in each energy bin, an isotropic power-law energy spectrum $\Phi_\nu \propto E_\nu^{-\gamma}$ for all flavors of neutrinos and anti-neutrinos, in agreement with theoretical expectations\ \cite{Anchordoqui:2013dnh} and IceCube findings\ \cite{Aartsen:2017sml}.  The value of $\gamma$ is obtained from a fit to data in each bin.  This makes our results robust against variations with energy of the spectral shape of astrophysical neutrinos, unlike \Ref\ \cite{Aartsen:2017kpd}, which assumed a single power law spanning the range 6.3--980 TeV.  We assume flavor equipartition, as expected from standard mixing\ \cite{Beacom:2003nh, Kashti:2005qa, Lipari:2007su, Mena:2014sja, 
Palomares-Ruiz:2015mka, Bustamante:2015waa, Vincent:2016nut} and in agreement with data\ \cite{Aartsen:2015ivb, Aartsen:2015knd}.
Because IceCube cannot distinguish neutrinos from anti-neutrinos, we can only extract a combination of their cross sections, each weighed by its corresponding flux.  We assume the likely case\ \cite{Murase:2013rfa, Murase:2015xka} of equal fluxes, coming, \eg, from proton-proton interactions\ \cite{Kelner:2006tc}. 


{\bf Assumptions.---}  Because data is scant, to reduce the number of free parameters to fit, we make three reasonable assumptions inspired on standard high-energy predictions.  With more data, they could be tested.

First, the rate of CC showers dominates over the rate of NC showers at any value of $E_\text{dep}$, based on the arguments above.  For simplicity, we adopt a constant $\langle y \rangle = 0.25$ for NC showers.  This assumption allows us to express the extracted cross section as a function of $E_\nu \approx E_\text{dep}$.

Second, CC cross sections dominate over NC cross sections.  We assume $\sigma_{\nu N}^\text{NC} = \sigma_{\nu N}^\text{CC} / 3$ and $\sigma_{\bar{\nu} N}^\text{NC} = \sigma_{\bar{\nu} N}^\text{CC} / 3$, following, \eg, \Ref\ \cite{Connolly:2011vc}.  This assumption allows us to fit only for CC cross sections.

Third, the ratio of $\bar{\nu} N$ to $\nu N$ cross sections is fixed in each bin.  Hence, when fitting, $\sigma_{\bar{\nu} N}^\text{CC} = \langle \sigma_{\bar{\nu} N}^\text{CC} / \sigma_{\nu N}^\text{CC} \rangle \cdot \sigma_{\nu N}^\text{CC}$, where $\langle \sigma_{\bar{\nu} N}^\text{CC} / \sigma_{\nu N}^\text{CC} \rangle$ is the average ratio in that bin predicted by \Ref\ \cite{CooperSarkar:2011pa} (see Table\ \ref{tab:fit_results}).  This assumption allows us to fit only for $\nu N$ cross sections. 

Thus, within each energy bin, we independently vary only the $\nu N$ CC cross section $\sigma_{\nu N}^\text{CC}$ and three nuisance parameters --- the number of showers due to atmospheric neutrinos $N_\text{sh}^\text{atm}$, the number of showers due to astrophysical neutrinos $N_\text{sh}^\text{ast}$, and the astrophysical spectral index $\gamma$.  To avoid introducing bias, we assume flat priors for all of them.  For each choice of values, we compare our test shower spectrum to the HESE shower spectrum via a likelihood.  To find the best-fit values of the parameters, we maximize the likelihood.  The Supplemental Material describes the statistical analysis in detail.


{\bf Results.---}  Table \ref{tab:fit_results} shows the extracted cross section, marginalized over the nuisance parameters.  Because $\sigma_{\nu N}$ and $\sigma_{\bar{\nu} N}$ are not independent in the fit, we present their average there and in Figs.\ \ref{fig:hese-costhz-energy} and \ref{fig:cross_sections_compare}.

\begin{table}[t!]
 \begin{ruledtabular}
  \caption{\label{tab:fit_results}Neutrino-nucleon charged-current inclusive cross sections, averaged between neutrinos ($\sigma_{\nu N}^\text{CC}$) and anti-neutrinos ($\sigma_{\bar{\nu} N}^\text{CC}$), extracted from 6 years of IceCube HESE showers.  To obtain these results, we fixed $\sigma_{\bar{\nu} N}^\text{CC} = \langle \sigma_{\bar{\nu} N}^\text{CC} / \sigma_{\nu N}^\text{CC} \rangle \cdot \sigma_{\nu N}^\text{CC}$ --- where $\langle \sigma_{\bar{\nu} N}^\text{CC} / \sigma_{\nu N}^\text{CC} \rangle$ is the average ratio of $\bar{\nu}$ to $\nu$ cross sections calculated using the standard prediction from \Ref\ \cite{CooperSarkar:2011pa} --- and $\sigma_{\nu N}^\text{NC} = \sigma_{\nu N}^\text{CC}/3$, $\sigma_{\bar{\nu} N}^\text{NC} = \sigma_{\bar{\nu} N}^\text{CC}/3$.  Uncertainties are $1\sigma$, statistical plus systematic, added in quadrature.}
  \centering
  \begin{tabular}{cccc}
   $E_\nu$ [TeV]         & $\langle E_\nu \rangle$ [TeV] & $\langle \sigma_{\bar{\nu} N}^\text{CC} / \sigma_{\nu N}^\text{CC} \rangle$ & $\log_{10}[ \frac{1}{2} ( \sigma_{\nu N}^\text{CC} + \sigma_{\bar{\nu} N}^\text{CC} ) / \text{cm}^2 ]$ \\
   \hline
   18--50                & 32                            & 0.752 & $-34.35 \pm 0.53$       \\                          
   50--100               & 75                            & 0.825 & $-33.80 \pm 0.67$       \\                        
   100--400              & 250                           & 0.888 & $-33.84 \pm 0.67$       \\
   400--2004             & 1202                          & 0.957 & $> -33.21 \; (1\sigma)$       
  \end{tabular}
 \end{ruledtabular}
\end{table}

Figure \ref{fig:cross_sections_compare} shows that, in each bin, results agree within 1$\sigma$ with widely used standard predictions.
The IceCube Collaboration has adopted the cross section from Cooper-Sarkar {\it et al.}\ \cite{CooperSarkar:2011pa}.
We include other calculations for comparison\ \cite{Gandhi:1995tf,Gandhi:1998ri,Connolly:2011vc,Block:2014kza,Arguelles:2015wba}.
All predictions are consistent with our measurements within errors.

Our results are consistent with the IceCube analysis\ \cite{Aartsen:2017kpd}, which found a cross section compatible with \Ref\ \cite{CooperSarkar:2011pa}.  Their smaller uncertainty is due to using $\sim$$10^4$ through-going muons.  However, by grouping all events in a single energy bin, their analysis did not probe the energy dependence of the cross section.  Like that analysis, our results are also consistent with standard cross-section predictions, but in narrower energy intervals.

Because the number of showers in each bin is small, statistical fluctuations weaken the interplay of downgoing versus upgoing showers described above.  To isolate the dominant statistical uncertainty, we minimized again the likelihood, this time keeping the nuisance parameters fixed at their best-fit values (see Table\ II in the Supplemental Material). 
The resulting uncertainty, attributed to statistics only, is $0.51$, $0.63$, and $0.62$ in the first three bins, where we have a measurement.  The systematic uncertainty, obtained by subtracting these values in quadrature from the total uncertainties in Table\ \ref{tab:fit_results} is $0.14$, $0.23$, and $0.25$ in each bin, slightly higher than in \Ref\ \cite{Aartsen:2017kpd}, due to a less detailed modeling of the detector.  While \Ref\ \cite{Aartsen:2017kpd} found comparable statistical and systematic uncertainties, we are presently dominated by statistics, since it uses an event sample that is smaller by a factor of $\sim$200. 

Nevertheless, our results disfavor new-physics models where the cross section rises sharply below 1 PeV\ \cite{AlvarezMuniz:2002ga, Anchordoqui:2003jr, Ahn:2003cza, Illana:2004qc, Hussain:2005dm, Anchordoqui:2006fn, Lykken:2007kp, Anchordoqui:2010hq, Marfatia:2015hva}.  Figure \ref{fig:cross_sections_compare} shows as example a model of TeV-scale gravity with large extra dimensions\ \cite{AlvarezMuniz:2001mk}.  While this model was disfavored by the LHC\ \cite{Aad:2011bw, Chatrchyan:2012taa}, we provide independent confirmation via a different channel.  More stringent tests of new-physics models, beyond the scope of this letter, should also consider the effect of modifications to the inelasticity distribution.


{\bf Limitations and improvements.---}  IceCube is sparsely instrumented and designed to detect the enormous light imprints made by high-energy neutrinos.  Except for high-energy muons, it cannot track individual particles or reconstruct $Q^2$ and $x$, unlike densely instrumented detectors.  Hence, we can only extract the cross section as a function of energy, integrated over other kinematic variables.  While we cannot extract individual PDFs, we can test their combination in the cross section.

Further, IceCube cannot distinguish if a particular shower was made in a CC or an NC interaction, and by a neutrino or an anti-neutrino.  The differences are too subtle to unequivocally identify them in individual showers\ \cite{Kowalski:2004qc}, but it might be possible to extract them statistically from a large enough data sample\ \cite{Li:2016kra}.

Lastly, we assumed that the astrophysical neutrino flux is isotropic\ \cite{Aartsen:2014cva, Aartsen:2015knd, Aartsen:2017ujz}.  Nevertheless, there are hints of a Galactic contribution\ \cite{Aartsen:2014cva, Neronov:2015osa, Aartsen:2015knd, Denton:2017csz}, with data allowing $< 14\%$ of the all-sky flux to come from the Galactic Plane\ \cite{Aartsen:2017ujz}.  If a Galactic flux is discovered, future cross-section analyses will need to acknowledge its anisotropy to avoid incorrectly attributing the distribution of arrival directions solely to in-Earth attenuation.


{\bf Summary and outlook.---}  We have extracted the energy dependence of the neutrino-nucleon cross section at energies beyond those available in man-made neutrino beams, making use of the high-energy reach of IceCube.  Our results are compatible with predictions based on nucleon structure extracted from scattering experiments at lower energies and disfavor extreme deviations that could stem from new physics in the TeV--PeV range.

It would be straightforward to repeat the present analysis using a larger HESE shower sample.  The proposed upgrade IceCube-Gen2\ \cite{Aartsen:2014njl} could have an event rate 5--7 times higher, thus reducing the impact of random fluctuations.  These showers could be combined with showers from the upcoming KM3NeT detector\ \cite{Adrian-Martinez:2016fdl}; their improved angular resolution of $\sim$2$^\circ$ above 50 TeV would allow for better estimates of in-Earth attenuation.  Starting tracks can also be considered, as long as one does not rely on predictions of the inelasticity distribution to reconstruct the parent neutrino energy.

An interesting possibility is to measure the inelasticity distribution\ \cite{Aartsen:2018vez}.  This can be done using starting tracks where the hadronic shower energy $E_\text{sh}$ and the outgoing muon energy $E_\mu$ are known individually, in order to reconstruct the inelasticity $y = (1+E_\mu/E_\text{sh})^{-1}$\ \cite{Anchordoqui:2006wc, Ribordy:2013set}.

At the EeV scale, differences between cross-section predictions increase.
Measuring $\sigma_{\nu N}$ at these energies would probe $x \sim m_W / E_\nu \lesssim 10^{-6}$, beyond the reach of laboratory scattering experiments.  This would prove instrumental in testing not only new physics, but also predictions of the potentially non-linear behavior of PDFs at low $x$, such as from BFKL theory\ \cite{Lipatov:1976zz, Kuraev:1976ge, Kuraev:1977fs, Balitsky:1978ic} and color-glass condensates\ \cite{Gelis:2010nm}; see, \eg, \Ref\ \cite{Henley:2005ms, Anchordoqui:2006ta, Albacete:2015zra}.  However, because the predicted neutrino flux at these energies, while uncertain, is smaller than at PeV, precision measurements of the cross section will likely be limited by statistics; see \Ref\ \cite{Bertone:2018dse} for details.  Nevertheless, large-volume neutrino detectors like ARA\ \cite{Allison:2011wk, Allison:2014kha, Allison:2015eky}, ARIANNA\ \cite{Barwick:2006tg, Barwick:2014pca}, GRAND\ \cite{Alvarez-Muniz:2018bhp}, and POEMMA\ \cite{Olinto:2017xbi}, might differentiate\ \cite{Anchordoqui:2004ma} between predictions, provided the event rate is high enough.


\section*{Acknowledgements}

MB is supported in part by NSF Grants PHY-1404311 and PHY-1714479, and by the Danmarks Grundforskningsfond Grant 1041811001.  AC is supported by NSF CAREER award 1255557.  This work used resources provided by the Ohio Supercomputer Center.  We thank Patrick Allison, John Beacom, Francis Halzen, Spencer Klein, Shirley Li, and Subir Sarkar for useful feedback and discussion.




%


\newpage

\appendix


\onecolumngrid


\section{Shower rates in IceCube}\label{app:shower_rates}

Below, $\nu$ stands for both neutrino and anti-neutrino, unless otherwise specified.


\subsection{Neutrino-induced events}

High-energy neutrinos deep-inelastic-scatter off nucleons in the Antarctic ice.  Charged-current (CC) interactions make charged leptons: $\nu_l + N \to l + X$ ($l = e, \mu, \tau$), where $N$ is either a neutron or a proton, and $X$ are final-state hadrons, mostly pions.  Neutral-current (NC) interactions make neutrinos: $\nu_l + N \to \nu_l + X$.  Outgoing hadrons receive a fraction $y$ of the initial neutrino energy --- known as the inelasticity --- while outgoing leptons receive $(1-y)$ of it.  Outgoing charged particles make Cherenkov light that is collected by IceCube photomultipliers buried in the ice.

The muon from a $\nu_\mu$ CC interaction leaves a track of Cherenkov light several kilometers long that, if it crosses the instrumented volume of IceCube, is typically clearly identifiable.  Muon tracks also come from the decay of taus, made by $\nu_\tau$ CC interactions, into muons, which occurs 17\% of the time.

All other final-state charged particles create particle showers localized around the interaction vertex.  A shower from final-state hadrons has a high neutron and pion content --- a hadronic shower.  In a NC interaction, this is the only shower, since the final-state neutrino exits the detector.  In a $\nu_e$ CC interaction, the electron creates an additional shower that consists mainly of electrons, positrons, and photons, and has low hadronic content --- an electromagnetic shower.  In a $\nu_\tau$ CC interaction, the tau decay creates a hadronic shower $66\%$ of the time and an electromagnetic shower $17\%$ of the time (the remaining $17\%$ of the time, the tau decays to a muon, which creates track).  IceCube does not resolve individually the lepton- and hadron-initiated showers; they are detected as a superposition.  Also, it is unable to distinguish between neutrinos and anti-neutrinos based only on total energy deposition.

Shower detection in IceCube is calorimetric: if the shower starts well within the detector --- like in HESE showers --- all of the shower energy is deposited in the ice, and most of it is collected by the photomultipliers.  The relation between the energy of the shower $E_\text{sh}$ and the energy of the incoming neutrino $E_\nu$ depends on the flavor of the neutrino and the type of the interaction.  In a $\nu_e$ CC interaction, all of the neutrino energy is given to the electromagnetic and hadronic showers.  In a $\nu_\tau$ CC interaction, about $30\%$ of the tau energy is lost to neutrinos at decay, after averaging over all decay channels.  In a NC interaction, on average, the shower energy is only $\langle y \rangle E_\nu$, where $\langle y \rangle$ is the average inelasticity.  Around $E_\nu = 1$ PeV, $\langle y \rangle \approx 0.25$ for neutrinos and anti-neutrinos, and for CC and NC interactions\ \cite{Gandhi:1995tf}.  In summary, the average fraction $f_{l,t}$ of neutrino energy carried by the shower in a $\nu_l$ or $\bar{\nu}_l$ interaction of type $t$ (CC or NC) is \cite{Bustamante:2016ciw}
\begin{equation}\label{equ:f_factors}
 f_{l,t} 
 \equiv
 \frac {E_\text{sh}} {E_\nu}
 \simeq
 \left\{\begin{array}{ll}
  1                                                                                      & \text{for } l = e \;\;\text{and}\;\; t = \text{CC} \\
  \left[ \langle y \rangle + 0.7 \left( 1 - \langle y \rangle \right) \right] \simeq 0.8 & \text{for } l = \tau \;\;\text{and}\;\; t = \text{CC} \\
  \langle y \rangle \simeq 0.25                                                          & \text{for } l = e, \mu, \tau \;\;\text{and}\;   \; t =\text{NC}
 \end{array}\right. \;.
\end{equation}
(See also \Ref\ \cite{Blum:2014ewa}, where different decay modes of the tau are treated separately.)  Since $f_{l,\text{NC}}$ is small, and since the atmospheric and astrophysical neutrino fluxes fall steeply with energy ($\propto E_\nu^{-\gamma}$), the NC contribution to the total shower rate is sub-dominant.


\subsection{Energy and angular spectrum of showers}

In the main text, we established that sensitivity to the neutrino-nucleon cross section comes from the attenuation of the neutrino flux as it propagates inside the Earth, which depends on neutrino energy and direction.  Therefore, to constrain the cross section, we need to compute the doubly differential spectrum --- in energy and arrival direction --- of showers in IceCube.  To do that, we extend the ``theorist's approach'' from \Refs\ \cite{Laha:2013eev, Bustamante:2016ciw} (see also \Ref\ \cite{Blum:2014ewa}) to account for the angular distribution:
\begin{equation}\label{equ:shower_rate_total_true}
 \frac { d^2N_\text{sh} } { d E_\text{sh} d \cos\theta_z }
 =
 \frac { d^2N_{\text{sh},e}^\text{CC} } { d E_\text{sh} d \cos\theta_z }
 + 
 \text{Br}_{\tau \to \text{sh}} \frac { d^2N_{\text{sh},\tau}^\text{CC} } { d E_\text{sh} d \cos\theta_z }
 +
 \sum_{l=e,\mu,\tau} \frac { d^2N_{\text{sh},l}^\text{NC} } { d E_\text{sh} d \cos\theta_z } \; ,
 \end{equation}
where $\theta_z$ is the zenith angle of the incoming neutrino (the normal to the South Pole is at $\theta_z = 0$), $\text{Br}_{\tau \to \text{sh}} = 0.83$ is the branching ratio of tau decays that make a shower, and
\begin{eqnarray}\label{equ:shower_rate_flavor_true}
 \frac { d^2N_{\text{sh},l}^\text{CC} } { d E_\text{sh} d \cos\theta_z } (E_\text{sh}, \cos \theta_z)
 \simeq
 - 2 \pi \rho_\text{ice} N_A V T \;
 &&
 \left\{
 \Phi_l (E_\nu)
 \sigma_{\nu N}^\text{CC} (E_\nu)
 e^{-\tau_{\nu N}(E_\nu, \theta_z)} 
 \right.
 \\
 &&
 \left.
 \left.
 + \,
 \Phi_{\bar{l}}(E_\nu) 
 \sigma_{\bar{\nu} N}^\text{CC} (E_\nu)
 e^{-\tau_{\bar{\nu} N}(E_\nu, \theta_z)}
 \right\}
 \right\vert_{E_\nu = E_\text{sh}/f_{l,\text{CC}}}
 \nonumber \;,
\end{eqnarray}
for showers initiated by CC interactions of a flux of $\nu_l$ ($\Phi_l$) and $\bar{\nu}_l$ ($\Phi_{\bar{l}}$).  On the right-hand side of \equ{shower_rate_flavor_true}, the neutrino energy is computed from the shower energy by means of \equ{f_factors}.  The number of nucleon targets inside the instrumented volume is $\rho_\text{ice} N_A V$, with $\rho_\text{ice} \approx 0.92$ g cm$^{-3}$ the density of ice, $N_A$ the Avogadro number, and $V \approx 1$ km$^3$ the volume of IceCube.  The expression for showers from NC interactions, $d^2N_{\text{sh},l}^\text{NC} / d E_\text{sh} / d \cos\theta_z$, is obtained from \equ{shower_rate_flavor_true} by changing $\sigma_{\nu N}^\text{CC} \to \sigma_{\nu N}^\text{NC}$, $\sigma_{\bar{\nu} N}^\text{CC} \to \sigma_{\bar{\nu} N}^\text{NC}$, and $f_{l,\text{CC}} \to f_{l,\text{NC}}$. 

To calculate the attenuation factors $e^{-\tau_{\nu N}}$ and $e^{-\tau_{\bar{\nu} N}}$ , consider an incoming flux of neutrinos with energy $E_\nu$ and zenith angle $\theta_z$.  Inside the Earth, which has approximate radius $R_\oplus = 6371$ km, the neutrinos travel a distance
\begin{equation}\label{equ:pathlength}
 D(\theta_z) = \sqrt{ \left( R_\oplus^2 - 2 R_\oplus d \right) \cos^2 \theta_z + 2 R_\oplus d } - \left( R_\oplus - d \right) \cos \theta_z
\end{equation}
before reaching IceCube, which is buried at a depth $d = 1.5$ km.  We compute the average Earth density $\langle \rho_\oplus(\theta_z) \rangle = (1/D(\theta_z)) \int_0^{D(\theta_z)} \rho_\oplus\left(x\right) dx$ encountered by the neutrino using the density profile $\rho_\oplus$ from the Preliminary Reference Earth Model (PREM)\ \cite{Dziewonski:1981xy,Gandhi:1995tf}.  (Variations between PREM and other Earth density models are at the level of 5\%, so they can be neglected given the size of the errors in our extracted cross sections.)  To a good approximation, Earth matter is isoscalar --- composed of equal numbers of neutrons and protons --- so the average nucleon mass is $m_N = (m_p + m_n)/2$.  Thus, the $\nu N$ interaction length (for any flavor) is
\begin{equation}\label{equ:int_length}
 L_{\nu N}(E_\nu, \theta_z)
 =
 \frac { m_N } { \langle \rho_\oplus (\theta_z) \rangle } 
 \left( \frac { 1 } { \sigma_{\nu N}^\text{CC}(E_\nu) + \sigma_{\nu N}^\text{NC}(E_\nu) } \right) \;,
\end{equation}
and, from this, the attenuation factor is
\begin{equation}
 e^{-\tau_{\nu N}( E_\nu, \theta_z )} 
 \equiv
 e^{ - D(\theta_z) / L_{\nu N}(E_\nu,\theta_z) } \;.
\end{equation}
For anti-neutrinos, the interaction length and attenuation factor have identical expressions, with $\nu \to \bar{\nu}$.

Figure \ref{fig:interaction_length} shows the interaction length as a function of zenith angle, computed, for illustration, using the standard prediction of the high-energy cross section from \Ref\ \cite{CooperSarkar:2011pa}.  There, we have separated the NC and CC interactions lengths, to illustrate the fact that the CC cross section is predicted to be $\sim 3$ times higher than the NC cross section.  

\setcounter{figure}{0}
\renewcommand{\thefigure}{A\arabic{figure}}
\begin{figure}[t!]
 \begin{minipage}{0.48\textwidth}  
   \centering
   \includegraphics[width=\textwidth]{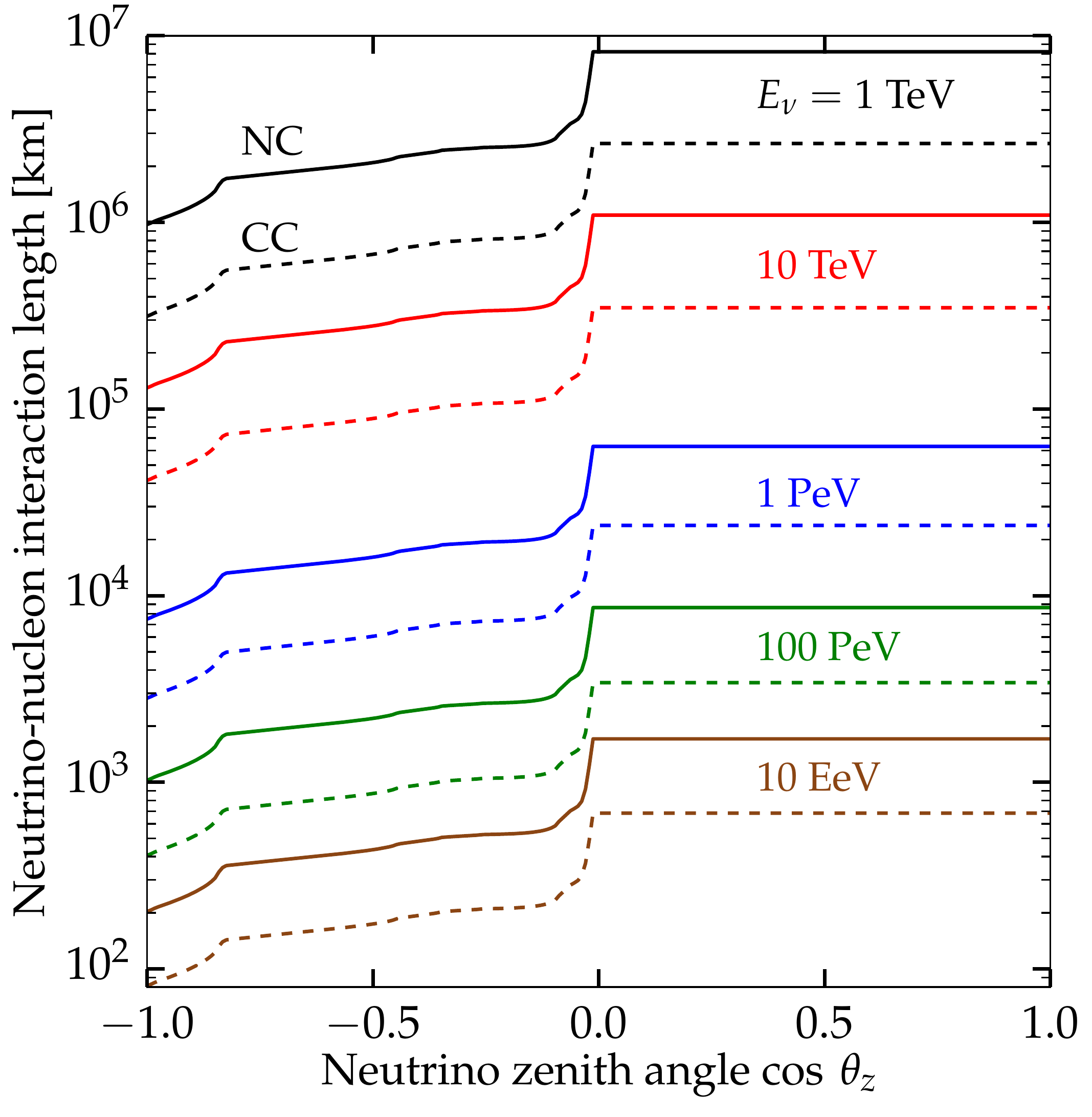}
   \caption{Neutrino-nucleon interaction length for neutrinos inside the Earth, as a function of zenith angle, for different neutrino energies.  We have separately calculated the length for neutral-current (NC, solid lines) and charged-current (CC, dashed lines) interactions.  The $\nu N$ cross sections are taken from \Ref\ \cite{CooperSarkar:2011pa}.  Interaction lengths for anti-neutrinos (not shown) are about $60\%$ higher, due to the smaller cross section.}
   \label{fig:interaction_length}
 \end{minipage}
 \hspace*{0.2cm}
 \begin{minipage}{0.48\textwidth}  
  \vspace*{-0.75cm}
  \centering
  \includegraphics[width=\textwidth]{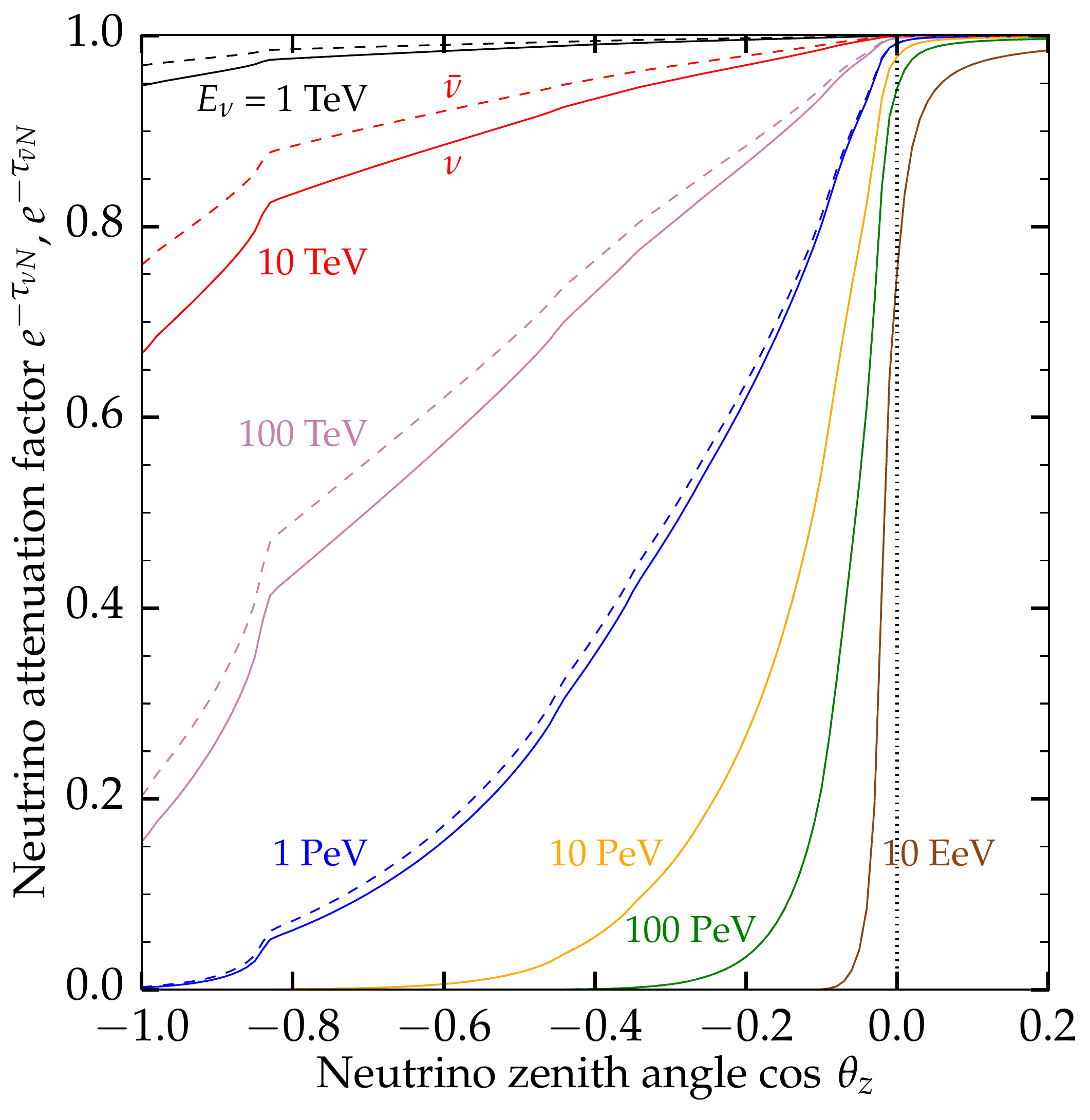}
  \vspace*{-0.47cm}
  \caption{In-Earth attenuation factors for neutrinos (solid lines) and anti-neutrinos (dashed lines), as a function of zenith angle, calculated for different neutrino energies and using the central values of the $\nu N$ cross sections from \Ref\ \cite{CooperSarkar:2011pa}.}
  \label{fig:attenuation_vs_costhz}
 \end{minipage}
\end{figure}

Figure \ref{fig:attenuation_vs_costhz} shows the corresponding attenuation factors.  Close to the horizon, attenuation is small ($e^{-\tau_{\nu N}} \approx 1$), except at very high energies, while above the horizon, attenuation is negligible at all energies.  Kinks on the curves reflect transitions between layers of different density inside the Earth\ \cite{Dziewonski:1981xy}.

The authors of \Refs\ \cite{Palomares-Ruiz:2015mka, Vincent:2017svp} performed a more comprehensive calculation of attenuation, treating different flavors of neutrinos and anti-neutrinos separately.  Our results are compatible with theirs, except for the inclusion of charged-current regenerations of $\nu_\tau$ and neutral-current regeneration of all flavors, which we have ignored since they affect the flux arriving at the detector only at the $\sim$10$\%$ level, which is unresolvable in the face of the large cross-section uncertainties we find. 

The contribution of atmospheric neutrinos to the HESE event rate is reduced by using the outer layer of PMTs as a veto.  When a contained event occurs, if the outer PMTs detect the passage of a muon that was made in the same atmospheric interaction as the neutrino responsible for the contained event, then the event is tagged as background.  Since the atmospheric neutrino flux falls faster with energy than the astrophysical flux, the probability that an atmospheric neutrino passes the veto falls with energy.  We have calculated the passing probability following \Refs\ \cite{Schonert:2008is, Gaisser:2014bja}, and multiplied \equ{shower_rate_flavor_true} by it when calculating the rate of showers due to atmospheric neutrinos.

In our analysis, we have not considered the fact that $\sim$30\% of IceCube contained tracks are mis-identified as showers\ \cite{Aartsen:2015ivb}, either because they deposit too little energy or because they occur too close to the edges of the detector.  In these events, because the shower is due mainly to the final-state hadrons, the deposited energy is small, \ie, $E_\text{dep} \approx y E_\nu$.  Hence, like NC showers, the contribution of mis-reconstructed muon tracks is sub-dominant.  Therefore, they should not significantly affect our ability to present the extracted cross sections as functions of $E_\nu \approx E_\text{dep}$.

At energies above 2 PeV --- beyond those available in the 6-year HESE sample --- we would need also to take into account showers created by $\bar{\nu}_e$ triggering the Glashow resonance\ \cite{Glashow:1960zz} on electrons ($\bar{\nu}_e + e \to W$), and the subsequent shower produced by the decay of the on-shell $W$.  At these energies, the shower rate due to neutrino-nucleon interactions is negligible, so any detection can be attributed to the Glashow resonance.  Thus, its eventual detection would single out the $\bar{\nu}_e$ flux and help break the degeneracy between neutrino and anti-neutrino cross sections.


\subsection{Astrophysical and atmospheric neutrino spectra}

For astrophysical neutrinos, we choose a power-law spectrum, in agreement with IceCube findings.  We assume equal proportions of each flavor in the flux, \ie, the flavor ratios are $( f_{e,\oplus} : f_{\mu,\oplus} : f_{\tau,\oplus} ) = ( \frac{1}{3} : \frac{1}{3} : \frac{1}{3} )$, which is compatible with IceCube results\ \cite{Aartsen:2015ivb,Aartsen:2015knd} and with theoretical predictions of standard flavor mixing\ \cite{Beacom:2003nh,Kashti:2005qa,Lipari:2007su,Mena:2014sja,Palomares-Ruiz:2015mka,Bustamante:2015waa,Vincent:2016nut}.  We also assume equal proportion of neutrinos and anti-neutrinos in the flux, which is expected from neutrino production in proton-proton interactions\ \cite{Kelner:2006tc} and, at high energies, in proton-photon interactions\ \cite{Kelner:2008ke, Hummer:2010vx}.  The spectrum of $\nu_l$ is
\begin{equation}\label{equ:spectrum_astro}
 \Phi_{l}^\text{ast} ( E_\nu )
 =
 \Phi_{\nu,0} \left( \frac {E_\nu} {100\ \text{TeV}} \right)^{-\gamma} \;,
\end{equation}
where $\Phi_{\nu,0}$ is the normalization per flavor of neutrino or anti-neutrino (in units of GeV$^{-1}$ cm$^{-2}$ s$^{-1}$ sr$^{-1}$) and $\gamma$ is the spectral index, common to all flavors, and to neutrinos and anti-neutrinos.  Equation (\ref{equ:spectrum_astro}) also describes the spectrum of $\bar{\nu}_l$.  Our analysis (Appendix \ref{app:statistical_analysis}) finds values of $\Phi_{\nu,0}$ (implicitly) and $\gamma$ inside each energy via a fit to IceCube data.

For conventional atmospheric neutrinos, created in the decays of pions and kaons, we use the recent calculation of $\nu_e$, $\bar{\nu}_e$, $\nu_\mu$, and $\bar{\nu}_\mu$ fluxes by Honda {\it et al.} from \Ref\ \cite{Honda:2015fha}.

We do not include prompt atmospheric neutrinos\ \cite{Bugaev:1989we, Gondolo:1995fq, Pasquali:1998ji, Enberg:2008te, Gaisser:2013ira, Bhattacharya:2015jpa, Garzelli:2015psa, Fedynitch:2015zbe, Halzen:2016pwl, Halzen:2016thi, Gaisser:2016obt, Bhattacharya:2016jce, Engel:2017wpa} in our analysis because recent searches have repeatedly failed to find them\ \cite{Aartsen:2013bka, Aartsen:2013jdh, Aartsen:2013eka, Aartsen:2014gkd, Aartsen:2014muf, Aartsen:2015xup, Aartsen:2015knd, Aartsen:2015rwa, Aartsen:2016xlq}.  However, in a full analysis of cross sections performed by the IceCube Collaboration, using more HESE data, the normalization of the prompt neutrino flux could be left as an additional free parameter to be determined by a fit, like it was done in \Ref\ \cite{Aartsen:2017kpd}.

In our calculations, we convert declination and right ascension to zenith angle using the {\tt astropy} package\ \cite{Robitaille:2013mpa}.  Since the Earth density profile model that we use is spherically symmetric\ \cite{Dziewonski:1981xy,Gandhi:1995tf}, the single zenith angle coordinate is sufficient to calculate the neutrino attenuation inside the Earth.


\subsection{Energy and angular resolution of the detector}

To compare our predicted shower spectra with the spectrum of observed HESE showers, we need to account for the energy resolution and angular resolution of the detector.  We do that by convolving the true spectrum, \equ{shower_rate_total_true}, with two functions that parametrize the detector resolution, \ie,
\begin{equation}\label{equ:shower_rate_total_rec}
 \frac { d^2N_\text{sh} } { d E_\text{dep} d \cos\theta_z }
 =
 \int dE_\text{sh} \int d\cos \theta_z^\prime
 \frac { d^2N_\text{sh} } { d E_\text{sh} d \cos\theta_z^\prime }
 R_E ( E_\text{sh}, E_\text{dep}, \sigma_E(E_\text{sh}) )
 R_\theta ( \cos \theta_z^\prime, \cos \theta_z, \sigma_{\cos \theta_z} ) \;,
\end{equation}
where the energy resolution function $R_E$ and the angular resolution function $R_\theta$ are Gaussians centered around the true values $E_\text{sh}$ and $\cos \theta_z^\prime$, respectively.

For the energy resolution function, we adopt\ \cite{Palomares-Ruiz:2015mka,Vincent:2016nut,Bustamante:2016ciw}
\begin{equation}
 R_E ( E_\text{sh}, E_\text{dep}, \sigma_E(E_\text{sh}) )
 =
 \frac { 1 } { \sqrt{ 2 \pi \sigma_E^2(E_\text{sh}) } } \exp\left[ - \frac { ( E_\text{sh}-E_\text{dep} )^2 } { 2 \sigma_E^2( E_\text{sh} ) } \right] \;,
\end{equation}
with $\sigma_E(E_\text{sh}) = 0.1 E_\text{sh}$, consistent with the value reported by IceCube\ \cite{Aartsen:2013vja}.

For the angular resolution function of showers, there is no conventional parametrization, to the best of our knowledge.  We adopt a resolution function in cosine of the zenith angle, \ie,
\begin{equation}
 R_\theta ( \cos \theta_z^\prime, \cos \theta_z, \sigma_{\cos \theta_z} )
 =
 \frac { 1 } { \sqrt{ 2 \pi \sigma_{\cos \theta_z}^2 } } \exp\left[ - \frac { ( \cos\theta_z^\prime - \cos\theta_z )^2 } { 2 \sigma_{\cos \theta_z}^2 } \right] \;.
\end{equation}
The dispersion $\sigma_{\cos \theta_z}$ is calculated, for a given value of $\theta_z = \arccos( \cos\theta_z )$, as the average between the upward and downward fluctuation in the cosine, \ie,
\begin{equation}
 \sigma_{\cos \theta_z}
 \equiv
 \frac{1}{2}
 \left[
 \left\vert  \cos( \theta_z + \sigma_{\theta_z} ) - \cos \theta_z \right\vert +
 \left\vert  \cos( \theta_z - \sigma_{\theta_z} ) - \cos \theta_z \right\vert
 \right] \;,
\end{equation}
where we choose a representative value of $\sigma_{\theta_z} = 15^\circ$ for the dispersion of the angle itself.  In reality, $\sigma_{\theta_z}$ is a function of deposited shower energy, with the resolution deteriorating towards low energies, as illustrated in \Fig\ 14 of \Ref\ \cite{Aartsen:2013vja}.  Our simplified choice captures the mean angular resolution of HESE showers without attempting to extract a proper resolution function from the aforementioned figure.


\subsection{Lower-energy IceCube contained events}

We avoid using lower-energy contained events --- Medium Energy Starting Events (MESE)\ \cite{Aartsen:2014muf, IC2yrMESEURL}, down to $E_\text{dep} \approx 1$~TeV --- due to the difficulty of correctly modeling how light absorption and scattering by ice distort the angular acceptance of IceCube\ \cite{Aartsen:2013rt,KleinPrivateComm}.
For HESE, these effects are mitigated due to their higher light yield (see, \eg, \Fig\ 3 in \Ref\ \cite{TalkXuICRC2017}), so we ignore them here without introducing large errors.


\section{Statistical analysis}\label{app:statistical_analysis}

To extract the neutrino-nucleon cross section, we compare our test shower spectra (see Appendix\ \ref{app:shower_rates}) with the observed spectrum of IceCube HESE\ \cite{Aartsen:2014gkd,Kopper:2015vzf,IC4yrHESEURL,TalkKopperICRC2017} showers.  We bin showers in $E_\text{dep}$ --- which, for showers, approximates $E_\nu$ (since NC showers are sub-dominant; see Appendix\ \ref{app:shower_rates}).  Because of limited data, we use only four bins: 18--50 TeV, 50--100 TeV, 100--400 TeV, and 400--2004 TeV.  The first three bins contain roughly the same number of events each (17--20), while the final bin contains only 3 events; Table\ \ref{tab:fit_results} contains the event numbers.  We perform a fit to shower data in each bin independently, as described below, employing a maximum likelihood method modeled after \Refs\ \cite{Palomares-Ruiz:2015mka,Vincent:2016nut}. 

In a bin containing $N_\text{sh}^\text{obs}$ observed showers, the likelihood is
\begin{equation}\label{equ:likelihood}
 \mathcal{L}
 =
 \frac {e^{ - ( N_\text{sh}^\text{atm} + N_\text{sh}^\text{ast} ) }} { N_\text{sh}^\text{obs}! }
 \prod_{i=1}^{N_\text{sh}^\text{obs}} \mathcal{L}_i \;,
\end{equation}
where $N_\text{sh}^\text{atm}$ is the number of showers due to atmospheric neutrinos and $N_\text{sh}^\text{ast}$ is the number of showers due to astrophysical neutrinos.  The partial likelihood $\mathcal{L}_i$ of the $i$-th shower in this bin captures the relative probability of the shower being from an atmospheric or an astrophysical neutrino.  It is computed as
\begin{equation}
 \mathcal{L}_i
 =
 N_\text{sh}^\text{atm} \, \mathcal{P}_i^\text{atm}
 +
 N_\text{sh}^\text{ast} \, \mathcal{P}_i^\text{ast} \;,
\end{equation}
where $\mathcal{P}_i^\text{atm}$ and $\mathcal{P}_i^\text{ast}$ are, respectively, the probability distribution for this shower to be generated by the atmospheric neutrino flux and by the astrophysical neutrino flux.  These are calculated as
\begin{eqnarray}
 \mathcal{P}_i^\text{atm}
 &=&
 \left(
 \int_{E_{\text{dep}}^\text{min}}^{E_{\text{dep}}^\text{max}} dE_\text{dep} \, \int_{-1}^{1} d \cos\theta_z \,  
 \frac{ d^2N_\text{sh}^\text{atm}} { dE_\text{dep} d\cos\theta_z }
 \right)^{-1}
 \left(
 \left. \frac{ d^2N_\text{sh}^\text{atm}} { dE_\text{dep} d\cos\theta_z } \right\vert_{ E_{\text{dep},i}, \cos \theta_{z,i} }
 \right)
 \;, \label{equ:prob_atm} \\
 \mathcal{P}_i^\text{ast}
 &=&
 \left(
 \int_{E_{\text{dep}}^\text{min}}^{E_{\text{dep}}^\text{max}} dE_\text{dep} \, \int_{-1}^{1} d \cos\theta_z \,  
 \frac{ d^2N_\text{sh}^\text{ast}} { dE_\text{dep} d\cos\theta_z }
 \right)^{-1}
 \left(
 \left. \frac{ d^2N_\text{sh}^\text{ast}} { dE_\text{dep} d\cos\theta_z } \right\vert_{ E_{\text{dep},i}, \cos \theta_{z,i} }
 \right)
 \;, \label{equ:prob_ast}
\end{eqnarray}
where $E_{\text{dep}}^\text{min}$ and $E_{\text{dep}}^\text{max}$ are the boundaries of the energy bin.  The double integrals represent the number of events in the energy bin, summed over all arrival directions.  The shower spectra $d^2 N_\text{sh}/dE_\text{dep}/d\cos\theta_z$ for atmospheric and astrophysical neutrinos are computed in Appendix \ref{app:shower_rates}.  Equations (\ref{equ:prob_atm}) and (\ref{equ:prob_ast}) depend on the four cross sections $\sigma_{\nu N}^\text{CC}$, $\sigma_{\bar{\nu} N}^\text{CC}$, $\sigma_{\nu N}^\text{NC}$, and $\sigma_{\bar{\nu} N}^\text{NC}$.  We assume the cross sections to be constant within each bin.  Equation (\ref{equ:prob_ast})  depends also on the astrophysical spectral index $\gamma$.

The full likelihood for one energy bin, \equ{likelihood}, is a function of 7 free parameters: $N_\text{sh}^\text{atm}$, $N_\text{sh}^\text{ast}$, $\sigma_{\nu N}^\text{CC}$, $\sigma_{\bar{\nu} N}^\text{CC}$, $\sigma_{\nu N}^\text{NC}$, $\sigma_{\bar{\nu} N}^\text{NC}$, and $\gamma$.  However, the three simplifying assumptions introduced in the main text reduce the number of free parameters to 4: $\sigma_{\nu N}^\text{CC}$, $N_\text{sh}^\text{atm}$, $N_\text{sh}^\text{ast}$, $\gamma$.  The latter three are treated as nuisance parameters.  

In each energy bin, we independently vary and fit for the four free parameters.  We choose a flat prior for all of the parameters.  To find the maximum of the likelihood, we use {\tt MultiNest}, an efficient implementation of the multinodal nested sampling algorithm\ \cite{Feroz:2007kg,Feroz:2008xx,Feroz:2013hea}, via the Python module {\tt PyMultiNest}\ \cite{Buchner:2014nha}.  The fitting procedure returns, in each bin, the best-fit value and uncertainty of $\sigma_{\nu N}^\text{CC}$.  From this, we calculate $\sigma_{\bar{\nu} N}^\text{CC} = \langle \sigma_{\bar{\nu} N}^\text{CC} / \sigma_{\nu N}^\text{CC} \rangle \cdot \sigma_{\nu N}^\text{CC}$; the values of $\langle \sigma_{\bar{\nu} N}^\text{CC} / \sigma_{\nu N}^\text{CC} \rangle$ are in Table\ \ref{tab:fit_results}.  Because, in our analysis, the $\nu N$ and $\bar{\nu} N$ cross sections are not independent, we present the average between them, $( \sigma_{\nu N}^\text{CC} + \sigma_{\bar{\nu} N}^\text{CC} ) / 2$.  Table\ \ref{tab:fit_results}, in the main text, shows the results, marginalized over the nuisance parameters.

Table\ \ref{tab:nuisance_results} shows, for completeness, the resulting values of the nuisance parameters after fitting.  In the main text, they are used to isolate the statistical and systematic uncertainties.

\begin{table}[t!]
 \begin{ruledtabular}
  \caption{\label{tab:nuisance_results}Best-fit values and $1\sigma$ uncertainties of the nuisance parameters in each energy bin: number of showers due to atmospheric neutrinos $N_\text{sh}^\text{atm}$, number of showers due to astrophysical neutrinos $N_\text{sh}^\text{ast}$, and astrophysical spectral index $\gamma$.}
  \centering
  \begin{tabular}{cccc}
   $E_\nu$ [TeV]         & $N_\text{sh}^\text{atm}$ & $N_\text{sh}^\text{ast}$ & $\gamma$ \\
   \hline
   18--50                & $4.2 \pm 4.9$            & $11.4 \pm 3.5$           & $2.38 \pm 0.31$       \\                          
   50--100               & $6.3 \pm 5.3$            & $11.7 \pm 4.5$           & $2.43 \pm 0.31$       \\                        
   100--400              & $6.4 \pm 6.0$            & $12.9 \pm 5.2$           & $2.49 \pm 0.31$       \\
   400--2004             & $1.2 \pm 1.0$            & $1.73 \pm 0.89$          & $2.37 \pm 0.32$       
  \end{tabular}
 \end{ruledtabular}
\end{table}


\end{document}